%% file: EXO-12-025_temp.tex
\pdfoutput=1

\documentclass[11pt,twoside,a4paper,cmspaper,final,collab]{cms-tdr}
\def\svnVersion{270720}\def\cmsCernNo{2013-037}\def\cmsCernDate{\today}\def\cmsMessage{Published in Physics Letters B as \href{http://dx.doi.org/10.1016/j.physletb.2014.11.026}{\doi{10.1016/j.physletb.2014.11.026.}}}

\begin{document}\cmsNoteHeader{EXO-12-025}

\hyphenation{had-ron-i-za-tion}
\hyphenation{cal-or-i-me-ter}
\hyphenation{de-vices}

\RCS$Revision: 270082 $
\RCS$HeadURL: svn+ssh://svn.cern.ch/reps/tdr2/papers/EXO-12-025/trunk/EXO-12-025.tex $
\RCS$Id: EXO-12-025.tex 270082 2014-12-02 19:43:00Z clint $
\newlength\cmsFigWidth
\ifthenelse{\boolean{cms@external}}{\setlength\cmsFigWidth{0.49\textwidth}}{\setlength\cmsFigWidth{0.55\textwidth}}
\newlength\cmsFigWidthTwo
\ifthenelse{\boolean{cms@external}}{\setlength\cmsFigWidthTwo{0.98\columnwidth}}{\setlength\cmsFigWidthTwo{0.48\textwidth}}
\ifthenelse{\boolean{cms@external}}{\providecommand{\cmsLeft}{top}}{\providecommand{\cmsLeft}{left}}
\ifthenelse{\boolean{cms@external}}{\providecommand{\cmsRight}{bottom}}{\providecommand{\cmsRight}{right}}
\newcommand{\usedLumiWZ}{19.5\fbinv\xspace}
\newcommand{\rhoT}{\ensuremath{\rho_\mathrm{TC}}\xspace}
\newcommand{\aT}{\ensuremath{\cmsSymbolFace{a}_\mathrm{TC}}\xspace}
\newcommand{\omegaT}{\ensuremath{\omega_\mathrm{TC}}\xspace}
\newcommand{\piT}{\ensuremath{\Pgp_\mathrm{TC}}\xspace}
\newcommand{\LT}{\ensuremath{L_\mathrm{T}}\xspace}
\providecommand{\FEWZ} {{\textsc{fewz}}\xspace}

\cmsNoteHeader{EXO-12-025} 
\title{Search for new resonances decaying via WZ to leptons in proton-proton collisions at $\sqrt{s}=8$\TeV}

\date{\today}

\abstract{
A search is performed in proton-proton collisions at $\sqrt{s}=8$\TeV for exotic particles decaying via WZ to fully leptonic final states with electrons, muons, and neutrinos. The data set corresponds to an integrated luminosity of 19.5\fbinv. No significant excess is observed above the expected standard model background. Upper bounds at 95\% confidence level are set on the production cross section of a $\PWpr$ boson as predicted by an extended gauge model, and on the $\PWpr \PW\cPZ$ coupling. The expected and observed mass limits for a $\PWpr$ boson, as predicted by this model, are 1.55 and 1.47\TeV, respectively. Stringent limits are also set in the context of low-scale technicolor models under a range of assumptions for the model parameters.
}

\hypersetup{%
pdfauthor={CMS Collaboration},%
pdftitle={Search for new resonances decaying via WZ to leptons in proton-proton collisions at sqrt(s)=8 TeV},%
pdfsubject={CMS},%
pdfkeywords={CMS, physics, technicolor}}

\maketitle 
\section{Introduction}
Many extensions of the standard model (SM) predict heavy charged gauge bosons, generically called {\PWpr}, that decay into a $\PW$$\cPZ$ boson pair~\cite{ssm1,ssm3,Birkedal:2004au,littleHiggs,Agashe:2008jb,Grojean:2011vu}. These extensions include models with extended gauge sectors, designed to achieve gauge coupling unification, and theories with extra spatial dimensions. There are also models in which the {\PWpr} couplings to SM fermions are suppressed, giving rise to a fermiophobic {\PWpr} with an enhanced coupling to \PW\ and \cPZ\ bosons~\cite{Beringer:1900zz, PhysRevD.78.031701}. Further, searches for {\PWpr} bosons that decay into \PW\cPZ\ pairs are complementary to searches in other decay channels~\cite{CMSWprimeLNu8TeV,CMSWprimeLNu7TeV,ATLASWprimeLNu7TeV,CMSDijet8TeV,ATLASDijet7TeV,Wprimetb2011, Aad:2012ej, Chatrchyan:2014koa, ATLAS:2014wra,Aad:2014aqa,Aad:2014xra}, many of which assume that the $\PWpr \to \PW\cPZ$ decay mode is suppressed. New \PW\cPZ\ resonances are also predicted in technicolor models of dynamical electroweak symmetry breaking~\cite{Susskind:1978ms, tc2, tc1}.

This Letter presents a search for exotic particles decaying to a $\PW\cPZ$ pair with $\PW\to\ell\cPgn$ and $\cPZ\to\ell\ell$, where $\ell$ is either an electron ($\Pe$) or a muon ($\Pgm$), $\cPgn$ denotes a neutrino, and the $\PW$ and $\cPZ$ bosons are allowed to decay to differently flavored leptons. The data were collected with the CMS experiment in proton-proton collisions at a center-of-mass energy $\sqrt{s}=8$\TeV at the CERN LHC and correspond to an integrated luminosity of 19.5\fbinv. Previous searches in this channel have been performed at the Tevatron~\cite{Abazov:2009eu} and at the LHC~\cite{Aad:2012vs, WprimeWZpaper,ATLASWprimeWZ8TeV}. The results have typically  been interpreted within the context of benchmark models such as an extended gauge model (EGM)~\cite{ssm3} and low-scale technicolor (LSTC) models~\cite{tc2, tc1}. The search conducted by CMS at $\sqrt{s}=7$\TeV~\cite{WprimeWZpaper} excluded EGM \PWpr bosons with masses below 1143\GeV and set stringent LSTC limits under a range of assumptions regarding model parameters. Complementary searches have also been conducted using the hadronic decays of the $\PW$ and $\cPZ$ bosons~\cite{PhysRevLett.104.241801,CMSWZtag7TeV,Chatrchyan:2012rva,ATLASDiboson7TeV,Khachatryan:2014hpa,Khachatryan:2014gha}.

The search at $\sqrt{s}=8$\TeV presented in this paper focuses on the fully leptonic channel, which is characterized by a pair of same-flavor, opposite-charge, isolated leptons with high transverse momentum (\pt) and an invariant mass consistent with that of the \cPZ\ boson. A third, high-\pt, isolated, charged lepton is also present, along with missing transverse momentum  associated with the neutrino. Background arises from other sources of three charged leptons, both genuine and misidentified. The primary background is the irreducible SM $\PW\cPZ$ production. Non-resonant events with no genuine \cPZ\ boson in the final state, such as top quark pair ($\text{t}\bar{\text{t}}$), multijet, \PW+jet, \PW$\gamma$+jet, and \PW\PW+jet production, are also considered. Only the first of these is expected to make a significant contribution. Also included are events with a genuine \cPZ\ boson decaying leptonically and a third misidentified or nonisolated lepton, such as \cPZ+jets (including \cPZ+heavy quarks) and \cPZ$\gamma$ processes. The final background category includes events with a genuine \cPZ\ boson decaying leptonically and a third genuine isolated lepton, dominated by $\cPZ\cPZ \to 4\ell$ decays in which one of the four leptons is undetected. Although irreducible, this contribution is not expected to be significant because of the small \cPZ\cPZ\ production cross section and dilepton decay branching fraction.

The search presented here follows the method applied in the previous analysis~\cite{WprimeWZpaper}, whereby a counting experiment is used to compare the number of observed events to the number of expected signal and background events. However, the new analysis benefits from the increase in center-of-mass energy to 8\TeV and also from improvements in lepton identification, particularly at high \pt. An increase in sensitivity is achieved at high {\PWpr} masses by using optimized isolation criteria that successfully take into account collimated leptons from highly boosted $\cPZ$ bosons. The larger center-of-mass energy alone increases the signal production cross section by roughly 45--70\% for {\PWpr} masses between $1000-1500$~\GeV, while the improved lepton isolation criteria contribute  a 50\% increase in signal efficiency over the same range. Additional improvements related to the optimization of selection criteria are also incorporated. Finally, as in the previous analysis~\cite{WprimeWZpaper}, the results are interpreted within the context of {\PWpr} bosons in extended gauge models  and vector particles in LSTC models.
\section{The CMS detector}
\label{sec:CMS}
The central feature of the CMS apparatus is a superconducting solenoid of 6\unit{m} internal diameter, providing a magnetic field of 3.8\unit{T}. Within the superconducting solenoid volume are a silicon pixel and strip tracker, a lead tungstate crystal electromagnetic calorimeter (ECAL), and a brass and scintillator hadron calorimeter (HCAL), each composed of a barrel and two endcap sections. Muons are measured in gas-ionization detectors embedded in the steel flux-return yoke outside the solenoid. Extensive forward calorimetry complements the coverage provided by the barrel and endcap detectors.

The ECAL energy resolution for electrons with transverse energy $\ET\approx45$\GeV from $\Z \to \Pe \Pe$ decays is better than 2\% in the central region of the ECAL barrel $(\abs{\eta} < 0.8)$, and is between 2\% and 5\% elsewhere. For low-bremsstrahlung electrons, where 94\% or more of their energy is contained within a $3 \times 3$ array of crystals, the energy resolution improves to 1.5\% for $\abs{\eta} < 0.8$~\cite{Chatrchyan:2013dga}.

Muons are measured in the pseudorapidity range $\abs{\eta}< 2.4$, with detection planes made using three technologies: drift tubes, cathode strip chambers, and resistive-plate chambers. Matching muons to tracks measured in the silicon tracker results in a \pt resolution between 1 and 5\%, for \pt values up to 1\TeV~\cite{Chatrchyan:2012xi}.

The particle-flow method~\cite{pf2,CMS-PAS-PFT-10-001} consists in reconstructing and identifying each single particle with an optimized combination of all subdetector information. The energy of photons is directly obtained from the ECAL measurement, corrected for zero-suppression effects. The energy of electrons is determined from a combination of the track momentum at the main interaction vertex, the corresponding ECAL cluster energy, and the energy sum of all bremsstrahlung photons attached to the track. The energy of muons is obtained from the corresponding track momentum.

A more detailed description of the CMS detector, together with a definition of the coordinate system used and the relevant kinematic variables, can be found elsewhere~\cite{:2008zzk}.

\section{Event simulation}

The \PYTHIA~6.426 event generator~\cite{pythia} and the CTEQ6L1~\cite{Pumplin:2002vw} parton distribution functions (PDFs) were  used for producing the EGM {\PWpr} and LSTC signal samples.  For the detailed simulation of the {\PWpr} samples,  \PYTHIA  was used for parton showering and hadronization with the Z2* tune~\cite{ref:z2star} for the underlying event simulation.  Cross sections are scaled to next-to-next-to-leading order (NNLO) values calculated with {\FEWZ} 2.0~\cite{Gavin:2010az}, and range from 27.96\unit{fb} to 0.33\unit{fb} for {\PWpr} masses between 1000 and 1500\GeV. Characteristic signal widths are between 100 and 168\GeV for the same mass range and are dominated by the detector resolution, since the natural widths vary from 33 to 54\GeV.

For the LSTC study we assume that the technihadrons $\rhoT$ and $\aT$ decay to $\PW\cPZ$. Since these two states are expected to be nearly mass-degenerate~\cite{tc1}, they would appear as a single feature in the $\PW\cPZ$ invariant mass spectrum, and we hereafter refer to them collectively as $\rhoT$. Since we do not expect a difference
in the kinematics between the \PWpr\ and LSTC signals, we use the \PWpr\ samples as the  default for the analysis, with the cross sections for LSTC as given by \PYTHIA. We consider the same relationship between the masses of the $\rhoT$ and $\piT$ technihadrons as used in Refs.~\cite{WprimeWZpaper} and~\cite{LesHouches09}, $M_{\piT}=\frac{3}{4}M_{\rhoT} - 25$\GeV, and also investigate the dependence of the results on the relative values of the $\rhoT$ and $\piT$ masses. The relationship between the masses significantly affects the
$\rhoT$ branching fractions~\cite{LesHouches09}. If $M_{\rhoT}<2M_{\piT}$, the decay
$\rhoT \to \PW\piT$ dominates, such that the branching fraction $\mathcal{B}(\rhoT \to \PW\cPZ)<10\%$. However, if the $\rhoT \to \PW\piT$ decay is kinematically inaccessible,
$\mathcal{B}(\rhoT \to \PW\cPZ)$ approaches 100\%. Following Ref.~\cite{LesHouches09} we also assume that the LSTC parameter $\sin\chi$ is equal to $1/3$. Changes in this parameter affect the branching fractions for decay into $\PW\cPZ$ and $\PW \piT$.

The \MADGRAPH~5.1~\cite{madgraph} and \POWHEG~1.1~\cite{Nason:2004rx, powheg, Alioli:2010xd, Melia:2011tj} generators are interfaced to \PYTHIA for parton showering, hadronization, and simulation of the underlying event. The SM $\PW\cPZ$ process, which is the dominant irreducible background, was generated with \MADGRAPH. The $\cPZ\cPZ$ process, which contributes when one of the leptons is either outside the detector acceptance or misreconstructed, was generated using \POWHEG. The instrumental backgrounds were produced using \MADGRAPH and include $\cPZ+$jets, \ttbar, $\cPZ\gamma$, $\PW\PW+$jets, and $\PW+$jets. The background contribution from QCD multijet events and from $\PW\gamma$ events was also studied in the simulation and found to be negligible. Next-to-leading order (NLO) cross sections are  used with the exception of the $\PW$+jets process, where the NNLO cross section is used. The {\PWpr} signal and SM processes used to estimate background were modeled using a full \GEANTfour~\cite{GEANT} simulation of the CMS detector.

For all the simulated samples, the additional proton-proton interactions in each beam crossing (pileup) were modeled by superimposing minimum bias interactions
(obtained using \PYTHIA with the Z2* tune) onto simulated events, with the multiplicity distribution matching the one observed in data.
\section{Object reconstruction and event selection }
The $\PW\cPZ\to 3\ell \,\nu$ decay is characterized by a pair of same-flavor, opposite-charge, high-\pt isolated leptons with an invariant mass consistent with a $\cPZ$ boson, a third, high-\pt isolated  lepton, and a significant amount of missing transverse momentum associated with the escaping neutrino. The analysis, therefore, relies on the reconstruction of three types of objects: electrons, muons, and \MET. The magnitude of the negative vector sum of transverse momenta of all reconstructed candidates is used to calculate \MET. The events are reconstructed using a particle-flow approach~\cite{pf2,CMS-PAS-PFT-10-001} and the details of the selection are provided below.

Candidate events are required to have at least three reconstructed leptons ($\Pe$, $\Pgm$) within the chosen detector acceptance of $\abs{\eta} <2.5\; (2.4)$ for electrons (muons). The events are selected online using a double-electron or double-muon trigger for final states with the $\cPZ$ boson decaying into electrons or muons, respectively.

The double-electron trigger requires two clusters in the ECAL with $\ET > 33$\GeV. The lateral spread in $\eta$ of the energy deposits comprising the cluster is required to be compatible with that of an electron. The trigger also requires that the sum of the energy detected in the HCAL in a cone of $\Delta R < 0.14$, where $\Delta R = \sqrt{\smash[b]{(\Delta \phi)^2 + (\Delta \eta)^2}}$, centered on the cluster, be no more than 15\%\;(10\%) of the cluster energy in the barrel (endcap) region of the ECAL. Finally, the clusters are matched in $\eta$ and $\phi$ to a track that includes hits in the pixel detector.

The double-muon trigger requires a \textit{global} muon with $\pt > 22$\GeV and a \textit{tracker} muon with $\pt > 8$\GeV. The global muon is reconstructed using an \textit{outside-in} approach whereby each muon candidate in the muon system is matched to a track reconstructed in the tracker and a global fit combining tracker and muon hits is performed~\cite{Chatrchyan:2012xi}. The tracker muon is reconstructed using an \textit{inside-out} approach in which all tracks that are considered as possible muon candidates are extrapolated out to the muon system. If at least one muon segment  matches the extrapolated track, it qualifies as a tracker muon. The trigger requirements described above have been changed from those in Ref.~\cite{WprimeWZpaper} wherein two global muons were required to pass the online selection. The new  requirements improve sensitivity for collimated muons from highly boosted $\cPZ$ bosons.

Simulated events are weighted according to trigger efficiencies measured, in both observed and simulated data, using the ``tag-and-probe'' technique~\cite{CMS:2011aa} with a large $\cPZ\to\ell\ell$ sample. In the electron channel, we apply a parametrization based on the turn-on curve measured with observed electrons and find trigger efficiencies to be above 99\%. Muon trigger efficiencies above the turn-on are typically measured to be above 90\% in observed events.  Scale factors are also applied to the simulated samples to account for differences between the observed and simulated trigger efficiencies. These are approximately unity for both the electron and muon channels.

Candidates for leptons from the $\PW$ and $\cPZ$ boson decays are also required to pass a series of identification and isolation criteria designed to reduce background from jets that are misidentified as leptons. Electron candidates are reconstructed from a collection of electromagnetic clusters with matched tracks. The electron momentum is obtained from a fit to the electron track using a Gaussian-sum filter algorithm~\cite{gsf} along its trajectory taking into account the possible emission of bremsstrahlung photons in the silicon tracker. We require $\pt>35\,(20)\GeV$ for the electrons from the $\cPZ\;(\PW)$ boson decay. We also require $\abs{\eta} < 2.5$ and exclude the barrel and endcap transition region ($ 1.444 < \abs{\eta} < 1.566$) as electron reconstruction in this region is not optimal. In comparison with the requirements imposed on electrons from the $\PW$ boson decays, a looser set of identification requirements, primarily based on the spatial matching between the track and the electromagnetic cluster, is imposed for the electrons from the $\cPZ$ boson decays. Electron candidates are also required to be isolated with particle-flow-based relative isolation, $I_\text{rel}$, less than 0.15, where $I_\text{rel}$ is defined as the sum of the transverse momenta of all neutral and charged reconstructed particle-flow candidates inside a cone of $\Delta R  < 0.3$ around the electron in $\eta$-$\phi$  space  divided by the \pt of the electron. The $I_\text{rel}$ computation includes an event-by-event correction applied to account for the effect of pileup~\cite{ele-iso-PU-corr}. Finally, if an electromagnetic cluster associated with a photon from internal bremsstrahlung in $\PW$ and $\cPZ$ boson decays happens to be closely aligned with a muon track, it may be misreconstructed as an electron.  In order to remove such instances of misreconstruction, electrons are rejected if they are within a cone of $\Delta R<0.01$ around a muon. Observed-to-simulated scale factors for these identification and isolation requirements, measured using tag-and-probe and parametrized as a function of electron \pt and $\abs{\eta}$, are applied as corrections to the simulated samples.

Global muon candidates are reconstructed using information from both the silicon tracker and the muon system. Candidates are required to have at least one muon chamber hit that is included in the global muon track fit and at least two matched segments in the muon system. We require muons with  $\abs{\eta} < 2.4$ and leading (sub-leading) muon $\pt > 25\; (10)$\GeV for the muons from the $\cPZ$ decay and $\pt > 20$\GeV for the muons from the $\PW$ decay. We also require $\delta \pt/\pt < 0.3$ for the track used for the momentum determination, where $\delta \pt$ is the uncertainty on the measured transverse momentum, and we eliminate cosmic ray background by requiring that the transverse impact parameter of the muon with respect to the primary vertex position be less than 2\unit{mm}. Particle-flow-based relative isolation, with pileup corrections applied~\cite{muon-iso-PU-corr}, is defined using a cone of size $\Delta R <0.4$ around the primary muon and is required to be less than 0.12. The above identification criteria are modified  for muons coming from the $\cPZ$ boson decay: one of the muons is allowed to be a tracker muon only and the requirement on the number of muon chamber hits is removed.  Additionally, the isolation variable for each muon is modified to remove the contribution of the other muon.  These modifications improve the signal efficiency and hence the overall sensitivity for high-mass \PWpr\ bosons. Simulated samples are corrected using observed-to-simulated scale factors that are parametrized as a function of muon $\abs{\eta}$.

Opposite-sign, same-flavor lepton pairs with invariant mass between 71 and 111\GeV, consistent with the $\cPZ$ boson mass,  are used to reconstruct $\cPZ$ boson candidates. In the case of more than one $\cPZ$ boson candidate, where the two candidates share a lepton, the candidate with the mass closest to the nominal $\cPZ$ boson mass~\cite{Beringer:1900zz} is selected. Events with two distinct $\cPZ$ boson candidates, where the candidates do not share a lepton, are rejected  in order to suppress the $\cPZ\cPZ$ background. The charge misidentification rate for the leptons considered in the analysis is very small and thus neglected.

A candidate for the charged lepton from the decay of a $\PW$ boson, in the following referred to as a $\PW$ lepton, is then selected out of the remaining leptons. When several candidates are found, the one with the highest \pt is selected.  Neutrinos from the leptonic $\PW$ boson decays escape from the detector without registering a signal and result in significant \MET in the event. In order to increase the purity of the selection of $\PW$ boson decays, the \MET in the event is required to be larger than 30\GeV. This requirement discriminates against both high-\pt jets misidentified as leptons and photon conversions, where the source of the misidentified jet or photon can come from $\cPZ$+jets or $\cPZ\gamma$ events, respectively.

In order to suppress events where final-state radiation produces additional leptons (via photon conversion) that are identified as the $\PW$ lepton, we apply two additional requirements on the event after the $\PW$ lepton selection. First, events with the trilepton invariant mass $m_{3\ell} <$ 120\GeV are rejected to remove events where $m_{3\ell}$ is close to the $\cPZ$ boson mass. Second, events where the $\Delta R$ between either lepton from the $\cPZ$ boson decay and the $\PW$ lepton is less than 0.3 are rejected.  This removes cases where the $\PW$ lepton candidate comes from a converted photon and is unlikely to occur in the boosted topology of a massive \PWpr\ boson decay.

\begin{figure}[htbp]
\centering
\includegraphics[width=\cmsFigWidthTwo]{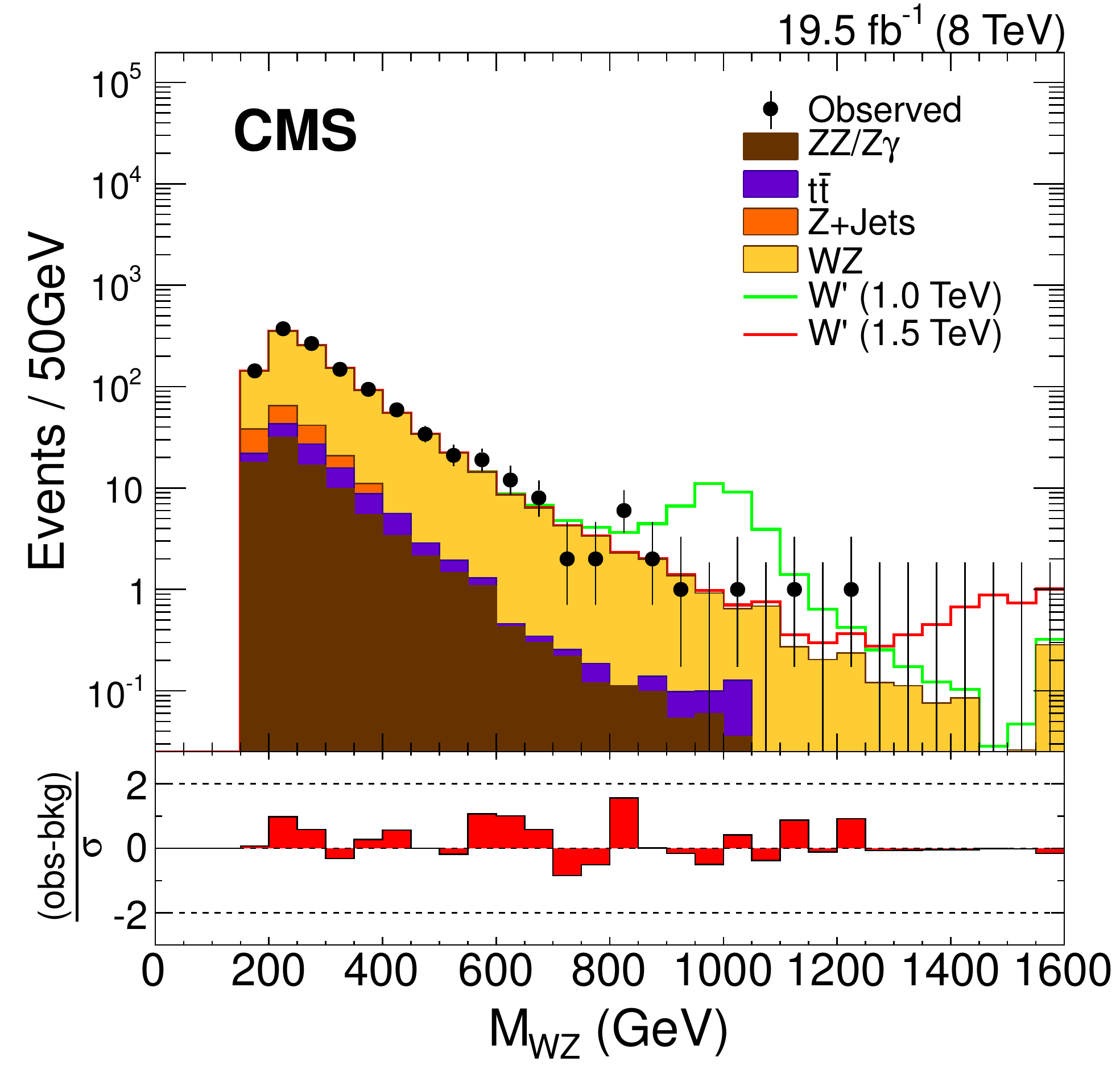}
\includegraphics[width=\cmsFigWidthTwo]{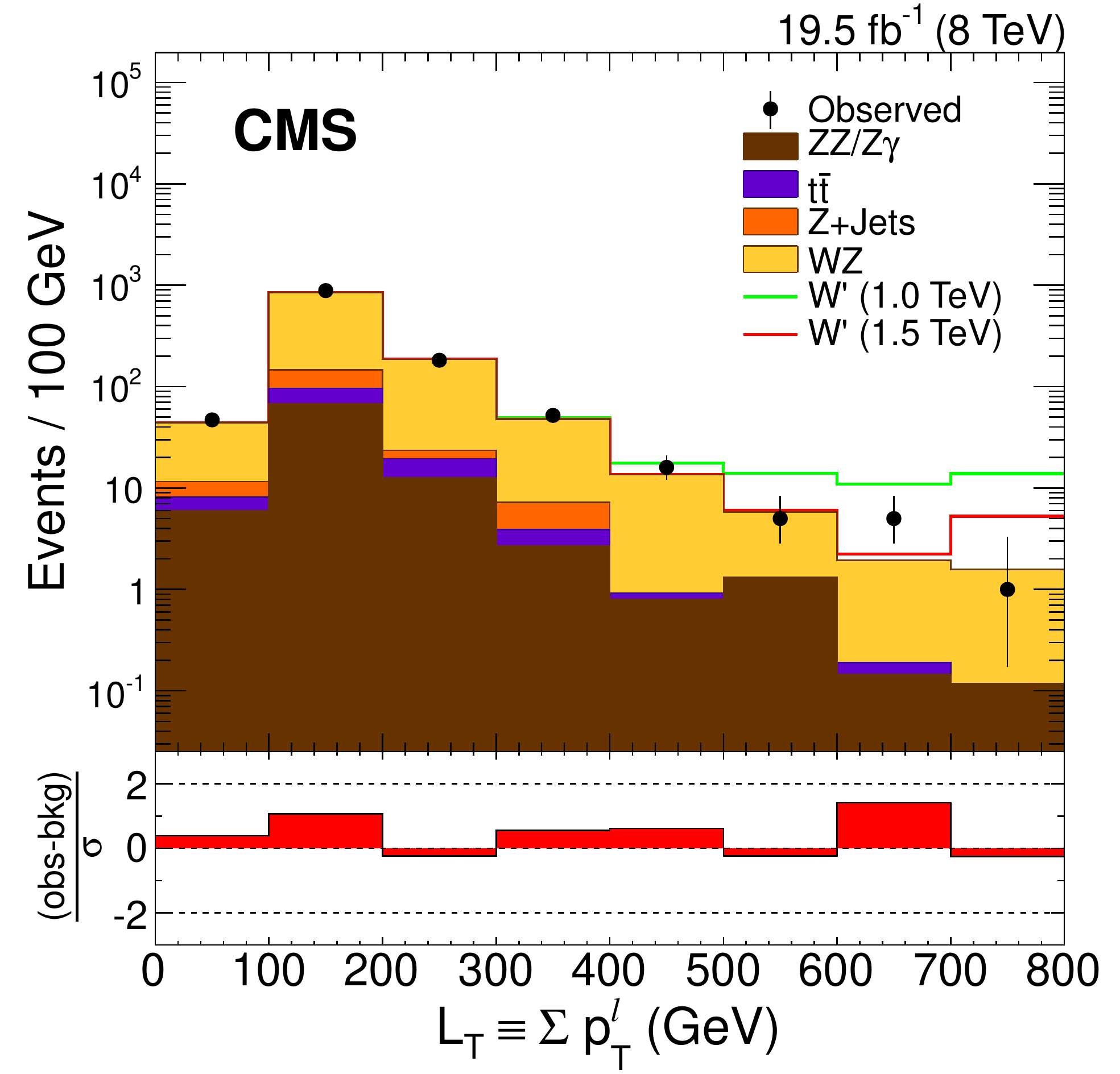}
\caption{The $\PW\cPZ$ invariant mass (\cmsLeft) and \LT (\cmsRight) distributions for the background, signal, and observed events  after
  the $\PW\cPZ$ candidate selection. The last bin includes overflow events. The $(\text{obs}-\text{bkg})/\sigma$ in the lower panel is defined as the difference between the number of observed events and the number of expected background events divided by the total statistical uncertainty.}
\label{fig:WZ-LT}

\end{figure}

After the $\PW$ and $\cPZ$ candidate selection, the two bosons are combined into a $\PW\cPZ$ candidate.  The invariant mass of this candidate cannot be determined uniquely since the longitudinal momentum of the neutrino is unknown. We follow the procedure used in the previous CMS analysis~\cite{WprimeWZpaper} and assume the $\PW$ boson to have its nominal mass, thereby constraining the value of the neutrino longitudinal momentum to one of the two solutions of a quadratic equation. Detector resolution effects can result in a reconstructed transverse mass larger than the invariant $\PW$ boson mass, $M_{\PW}$, leading to complex solutions for the neutrino longitudinal momentum. In these cases, a real solution is recovered by setting $M_{\PW}$ equal to the measured transverse mass. This results in two identical solutions for the neutrino longitudinal momentum. In simulated events with two distinct, real solutions, the smaller-magnitude solution was found to be correct in approximately 70\% of the cases, and this solution was therefore chosen for all such events.  Figure~\ref{fig:WZ-LT} (\cmsLeft) shows the $\PW\cPZ$ invariant mass distributions, after the $\PW\cPZ$-candidate selection, for signal, background, and observed events. At this point, the irreducible $\PW\cPZ$ process dominates the background contribution, making up $\sim$85\% of the total number of expected background events.

In order to further suppress SM background events, we apply two additional
selection requirements. The first is a requirement on \LT, the scalar sum of the charged leptons' transverse
momenta, shown in Fig.~\ref{fig:WZ-LT} (\cmsRight). The second is a requirement on the mass of the $\PW\cPZ$ system. The thresholds for these selection criteria are varied simultaneously at 100\GeV mass spacing for the $\PW\cPZ$ invariant mass and optimized for the best expected limit on the $\PWpr$ production. These optimal values are then plotted as a function of the $\PW\cPZ$ mass and an analytic function is fit to the resulting distribution. For the mass-window requirement, two regimes of linear behavior are observed: for masses less than 1200\GeV, a narrow mass window is optimal in order to reject as much background as possible. Above 1200\GeV, the background ceases to contribute significantly and it is better to have a large mass window. The \LT requirement exhibits a linear relationship: as the mass increases, it is optimal to require a larger \LT, until around 1000\GeV, at which point having \LT greater than 500\GeV is sufficient. These mass windows and \LT requirements are summarized in Table~\ref{tab:MassWindows-combined}.

\begin{table*}[htbp]
\topcaption{Minimum $\LT$ requirements and search windows for each EGM \PWpr\ mass point along
with the number of expected background events ($N_{\text{bkg}}$), observed events ($N_{\text{obs}}$), expected \PWpr\ signal events ($N_{\mathrm{sig}}$), and the product of the signal efficiency and acceptance ($\varepsilon_{\text{sig}} \times \text{Acc.}$).  The indicated uncertainties are statistical only.}
\centering \begin{tabular}{c*{6}{c}} \hline
\shortstack{\PWpr Mass \\ (\GeVns)} & \shortstack{$\LT$ \\ (\GeVns)} & \shortstack{$M_{\PW\cPZ}$ Window \\ (\GeVns)} & $N_{\text{bkg}}$ & $N_{\text{obs}}$ & $N_{\text{sig}}$ & \shortstack{$\varepsilon_{\text{sig}} \times \text{Acc.}$ \\ (\%)} \\
\hline
 170 & 110 & 163--177 & 9.0 $\pm$ 0.3 & 8 & 18 $\pm$ 1 & 1.33 $\pm$ 0.09 \\
 180 & 115 & 172--188 & 38 $\pm$ 2 & 49 & 140 $\pm$ 7 & 1.97 $\pm$ 0.09  \\
 190 & 120 & 181--199 & 62 $\pm$ 1 & 76 & 371 $\pm$ 14 & 2.6 $\pm$ 0.1  \\
 200 & 125 & 190--210 & 81 $\pm$ 4 & 86 & 610 $\pm$ 20 & 3.2 $\pm$ 0.1  \\
 210 & 130 & 199--221 & 86 $\pm$ 3 & 101 & 786 $\pm$ 23 & 3.9 $\pm$ 0.1  \\
 220 & 135 & 208--232 & 91 $\pm$ 3 & 84 & 896 $\pm$ 24 & 4.5 $\pm$ 0.1  \\
 230 & 140 & 217--243 & 92 $\pm$ 4 & 80 & 977 $\pm$ 25 & 5.2 $\pm$ 0.1  \\
 240 & 145 & 226--254 & 91 $\pm$ 4 & 84 & 1011 $\pm$ 24 & 5.8 $\pm$ 0.1  \\
 250 & 150 & 235--265 & 82 $\pm$ 1 & 85 & 1021 $\pm$ 23 & 6.4 $\pm$ 0.1  \\
 275 & 162 & 258--292 & 73 $\pm$ 3 & 85 & 970 $\pm$ 20 & 8.0 $\pm$ 0.2  \\
 300 & 175 & 280--320 & 60 $\pm$ 1 & 74 & 858 $\pm$ 16 & 9.6 $\pm$ 0.2  \\
 325 & 188 & 302--348 & 56 $\pm$ 3 & 53 & 792 $\pm$ 13 & 11.8 $\pm$ 0.2  \\
 350 & 200 & 325--375 & 48 $\pm$ 3 & 37 & 699 $\pm$ 10 & 13.9 $\pm$ 0.2  \\
 400 & 225 & 370--430 & 32 $\pm$ 1 & 40 & 542 $\pm$ 7 & 18.1 $\pm$ 0.2  \\
 450 & 250 & 415--485 & 23.1 $\pm$ 0.8 & 26 & 399 $\pm$ 5 & 21.5 $\pm$ 0.2  \\
 500 & 275 & 460--540 & 16.6 $\pm$ 0.5 & 13 & 297 $\pm$ 3 & 24.8 $\pm$ 0.3  \\
 550 & 300 & 505--595 & 13.2 $\pm$ 0.6 & 14 & 220 $\pm$ 2 & 27.6 $\pm$ 0.3  \\
 600 & 325 & 550--650 & 10.0 $\pm$ 0.5 & 10 & 167 $\pm$ 2 & 30.4 $\pm$ 0.3  \\
 700 & 375 & 640--760 & 4.7 $\pm$ 0.2 & 4 & 96.9 $\pm$ 0.8 & 34.3 $\pm$ 0.3 \\
 800 & 425 & 730--870 & 2.8 $\pm$ 0.2 & 5 & 56.5 $\pm$ 0.5 & 36.5 $\pm$ 0.3 \\
 900 & 475 & 820--980 & 2.1 $\pm$ 0.2 & 4 & 35.0 $\pm$ 0.3 & 38.6 $\pm$ 0.3  \\
 1000 & 500 & 910--1090 & 1.4 $\pm$ 0.1 & 0 & 23.7 $\pm$ 0.2 & 43.3 $\pm$ 0.3  \\
 1100 & 500 & 1000--1200 & 0.8 $\pm$ 0.1 & 0 & 15.9 $\pm$ 0.1 & 46.8 $\pm$ 0.3  \\
 1200 & 500 & 1080--1320 & 0.58 $\pm$ 0.08 & 1 & 10.77 $\pm$ 0.07 & 49.1 $\pm$ 0.3  \\
 1300 & 500 & 1108--1492 & 0.56 $\pm$ 0.08 & 1 & 8.20 $\pm$ 0.04 & 56.1 $\pm$ 0.3  \\
 1400 & 500 & 1135--1665 & 0.60 $\pm$ 0.08 & 1 & 5.64 $\pm$ 0.03 & 57.3 $\pm$ 0.3  \\
 1500 & 500 & 1162--1838 & 0.57 $\pm$ 0.08 & 1 & 3.76 $\pm$ 0.02 & 57.5 $\pm$ 0.3  \\
 1600 & 500 & 1190--2010 & 0.56 $\pm$ 0.08 & 1 & 2.56 $\pm$ 0.01 & 57.7 $\pm$ 0.3  \\
 1700 & 500 & 1218--2182 & 0.50 $\pm$ 0.08 & 1 & 1.782 $\pm$ 0.009 & 57.6 $\pm$ 0.3  \\
 1800 & 500 & 1245--2355 & 0.44 $\pm$ 0.07 & 1 & 1.255 $\pm$ 0.007 & 58.0 $\pm$ 0.3  \\
 1900 & 500 & 1272--2528 & 0.39 $\pm$ 0.07 & 0 & 0.844 $\pm$ 0.005 & 55.0 $\pm$ 0.3  \\
 2000 & 500 & 1300--2700 & 0.36 $\pm$ 0.07 & 0 & 0.595 $\pm$ 0.003 & 54.7 $\pm$ 0.3  \\
\hline
\end{tabular}
\label{tab:MassWindows-combined}
\end{table*}

\section{Systematic uncertainties}
\label{sec:systematics}

Systematic uncertainties affecting the analysis can be grouped into four categories. In the first group we include uncertainties that are determined from simulation.
These include uncertainties in the lepton and \MET energy scales and resolution, as well as uncertainties in the PDFs. Following the recommendations of the PDF4LHC group~\cite{Alekhin:2011sk,PDF4LHC}, PDF and $\alpha_s$ variations of the MSTW2008~\cite{Martin:2009iq}, CT10~\cite{Nadolsky:2008zw}, and NNPDF2.0~\cite{Ball:2010de} PDF sets are taken into account and their impact on the $\PW\cPZ$ cross section estimated. Signal PDF uncertainties are taken into account only to derive uncertainty bands around the signal cross sections, as shown in Fig.~\ref{fig:LimitVsMass}, and do not impact the central limit.  An uncertainty associated with the simulation of pileup  is also taken into account.

The second group includes the systematic uncertainties affecting the observed-to-simulated scale factors for the efficiencies of the trigger, reconstruction, and identification requirements. These efficiencies are derived from  tag-and-probe studies, and the uncertainty in the ratio of the efficiencies is typically taken as the systematic uncertainty. For the $\cPZ \to \Pe\Pe$ channel, we assign a 2\% uncertainty related to the trigger scale factors, another 2\% to account for the difference between the observed and simulated reconstruction efficiencies, and an additional 1\% uncertainty related to the electron identification and isolation scale factors. For the $\cPZ \to \mu\mu$ channel,  we assign a 5\% uncertainty related to the trigger and another 2\% uncertainty due to the differences in the observed and simulated efficiencies of muon reconstruction. An additional 3\% uncertainty is assigned  to the muon identification and isolation scale factors to cover potential differences related to the boosted topology of the signal.

The third category comprises uncertainties in the background yield. These are dominated by the theoretical uncertainties associated with the $\PW\cPZ$ background. We consider  contributions coming from uncertainties related to the choice of PDF (described above), renormalization and factorization scales, and the SM $\PW\cPZ$ production modeling in \MADGRAPH. Scale uncertainties were determined by studying the variation of the cross section in the same
phase space of the analysis by varying the renormalization and factorization scales by a factor of two upwards and downwards with respect to their
nominal values. The largest observed variation is taken as a systematic uncertainty. This procedure results in uncertainties of 5\% for $\PW\cPZ$  masses up to
500\GeV and up to 15\% from 600\GeV to 2\TeV. As the \MADGRAPH sample used for simulating the $\PW\cPZ$ background contains explicit production of additional jets at matrix-element level, it provides a reasonable description of the process. The prediction is thus only
rescaled with a global factor to the total NLO cross section computed
with \MCFM 6.6~\cite{Campbell:2010ff}. To estimate uncertainties related to remaining modeling differences between the spectra predicted by
\MADGRAPH and true NLO predictions, we studied the ratio of the
$\PW\cPZ$ cross section in the phase space defined by the analysis selection criteria (for
each mass point) to the inclusive $\PW\cPZ$ cross section. We compared this ratio between \MADGRAPH and \MCFM and found differences of the order of 5\% for $\PW\cPZ$ masses up to
1\TeV, and of the order of 30\% between 1 and 2\TeV. These differences are taken as additional systematic uncertainties in the SM $\PW\cPZ$ background. For other background processes, the cross sections are varied by amounts estimated for the phase space relevant for this analysis as follows: $\cPZ\cPZ$ and $\cPZ$+jets by 30\%, $\ttbar$ by 15\%, and $\cPZ\gamma$ by 50\%.

Finally, an additional uncertainty of 2.6\% due to the measurement of the integrated luminosity is included~\cite{LUMIPAS}. Table~\ref{tab:sys-summary} presents a summary of the above systematic uncertainties.

\begin{table*}[htb]
  \topcaption{Summary of systematic uncertainties. Values are given for the impact on signal and background event yields. When the value of the uncertainty differs between the different decay modes of the $\PW$ and $\cPZ$ bosons and/or between different \PWpr masses considered, a range is quoted in order to provide an idea of the magnitude of the uncertainty, \ie its impact.}
  \label{tab:sys-summary}
  \centering
  \begin{tabular}{l c  c }
    \hline
    Systematic Uncertainty & Signal Impact & Background Impact \\
    \hline
    \MET Resolution \& Scale & 1--3\% & 1--23\% \\
    Muon \pt Resolution & 1--3\% & 0.5--5\% \\
    Muon \pt Scale & 1--2\% & 1--22\% \\
    Electron Energy Scale \& Resolution & 0.5--2\% & 1.5--12\% \\
    Pileup & 0.1--0.8\% & 0.5--5\% \\
    \hline
    Electron Trigger Efficiency & 2\% & 2\% \\
    Electron Reconstruction Efficiency & 2\% & 2\% \\
    Electron ID \& Isolation Efficiencies & 1\%  &  1\% \\
    Muon Trigger Efficiency & 5\%  & 5\%  \\
    Muon Reconstruction Efficiency & 2\% & 2\%  \\
    Muon ID \& Isolation Efficiencies & 3\%  & 3\%   \\
    \hline
    $\cPZ$+jets & --- & 30\% \\
    $\ttbar$ & --- & 15\% \\
    $\cPZ\gamma$ & --- & 50\% \\
    $\cPZ\cPZ$ & --- & 30\% \\
    $\PW\cPZ$ PDF  &--- & 5--10\% \\
    $\PW\cPZ$ Scale  &--- & 5--15\% \\
    $\PW\cPZ$ \MADGRAPH Modeling &--- & 5--30\% \\
    \hline
    Luminosity & 2.6\% & 2.6\% \\
    \hline
\end{tabular}
\end{table*}

\section{Results}
\label{sec:Results}

As shown in Fig.~\ref{fig:WZ-LT}, the data are compatible with the expected SM background and no significant excess is observed. Exclusion limits on the production cross section $\sigma (\Pp\Pp
\to \PWpr/\rhoT \to \PW\cPZ) \times
 \mathcal{B}(\PW\cPZ \to 3 \ell \nu)$  are determined using a counting experiment and comparing the number of observed events
to the number of expected signal and background
events.   The limits are calculated at 95\% confidence level (CL) by employing the  {\textsc{RooStats}}~\cite{roostats} implementation of Bayesian statistics~\cite{Beringer:1900zz} and assuming a flat prior for the signal production cross section. Systematic uncertainties, other than signal PDF uncertainties, are represented by nuisance parameters.
The results for the number of observed and expected background and signal events at different $\PWpr$ masses, along with the efficiency times acceptance, are given in Table~\ref{tab:MassWindows-combined}.

\begin{figure}[htb]
\centering
\includegraphics[width=\cmsFigWidth]{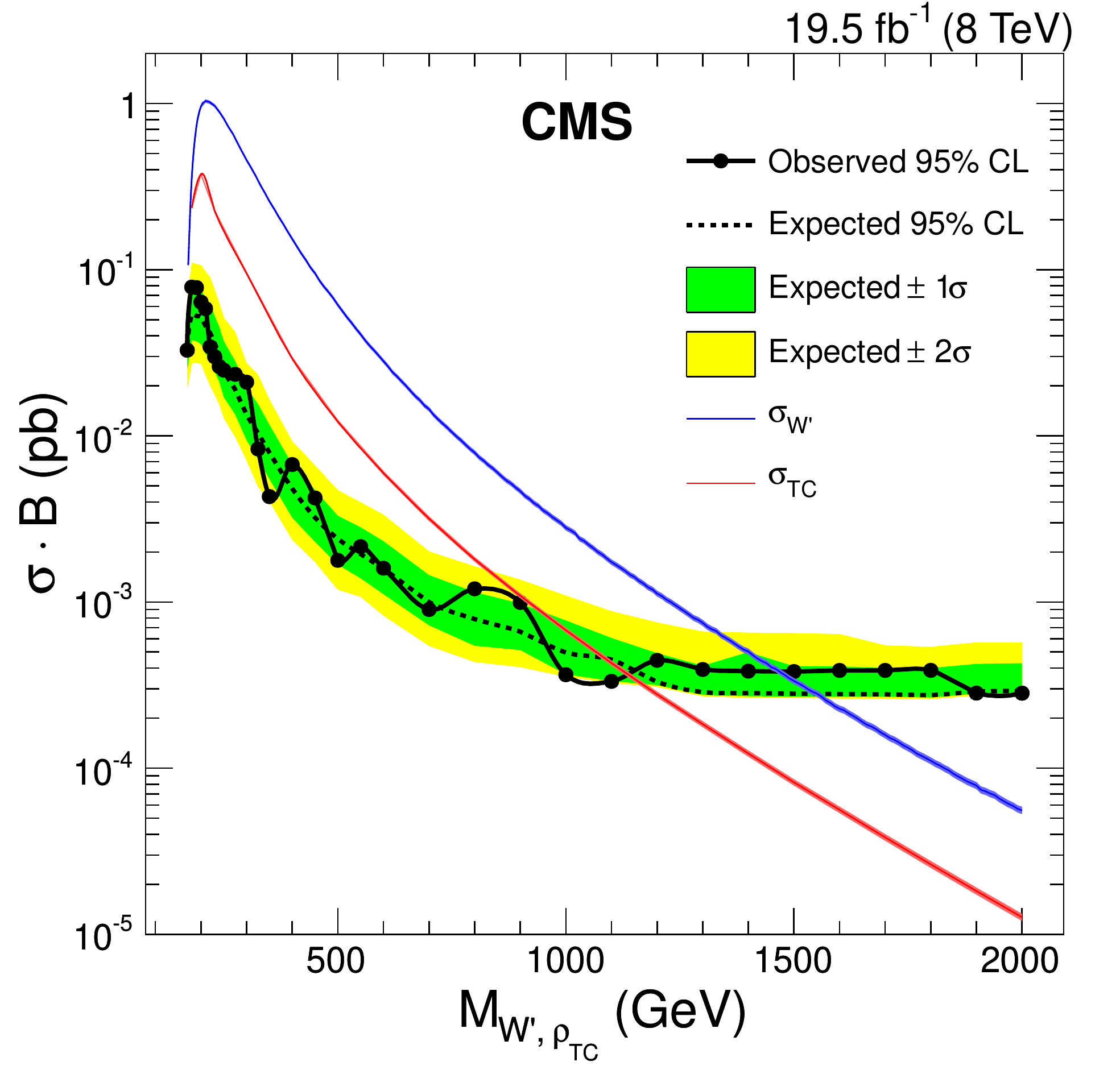}
 \caption{Limits at 95\% CL on $\sigma \times \mathcal{B}(\PWpr \to
   3 \ell \nu)$ as a function of the mass of the EGM $\PWpr$ (blue) and $\rhoT$ (red), along with the
1\,$\sigma$ and 2\,$\sigma$ combined statistical and systematic uncertainties indicated by the green (dark) and  yellow (light) band, respectively. The theoretical cross sections include a mass-dependent NNLO K-factor. The thickness of the theory lines represents the PDF uncertainty associated with the signal cross sections. The predicted cross sections for $\rhoT$ assume that $M_{\piT}=\frac{3}{4}M_{\rhoT} - 25$\GeV and that the LSTC parameter $\sin\chi = 1/3$.}
 \label{fig:LimitVsMass}

 \end{figure}

\begin{figure}[htb]
 \centering
 \includegraphics[width=\cmsFigWidth]{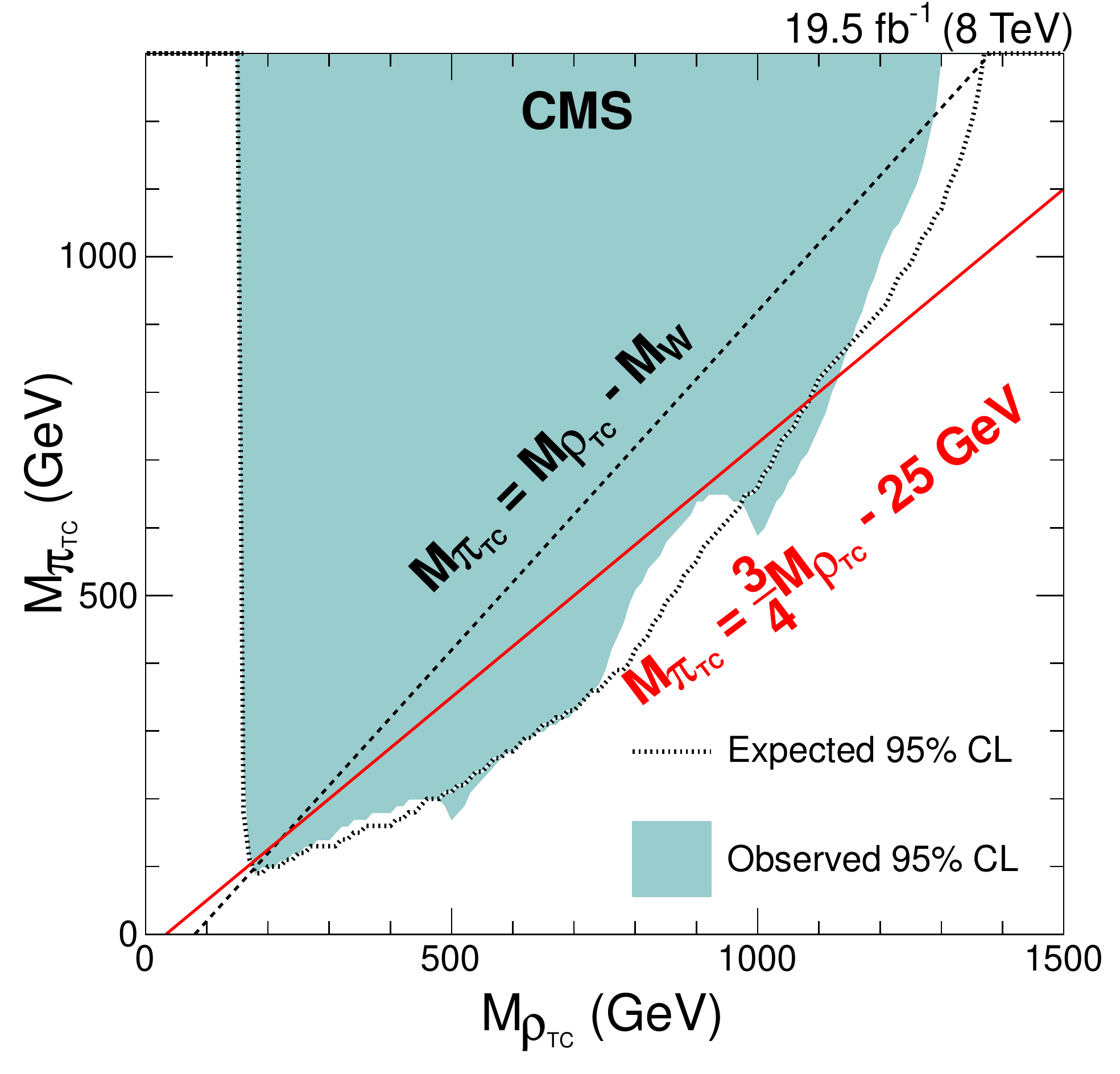}
 \caption{Two-dimensional exclusion limit at 95\% CL for the LSTC model as a function of the $\rhoT$ and $\piT$ masses.}
 \label{fig:TC2Dlimit}

 \end{figure}

The expected (observed) lower limit on the mass of the \PWpr\ boson is 1.55\,(1.47)\TeV in the EGM. For LSTC, with the chosen parameters $M_{\piT}=\frac{3}{4}M_{\rhoT} - 25\GeV$, the expected and observed $\rhoT$ mass limits are 1.09 and 1.14\TeV, respectively. For each of the above cases the lower bound on the exclusion limit is 0.17\TeV. Figure~\ref{fig:LimitVsMass} shows these limits as a function of the mass of the EGM \PWpr\ boson and the $\rhoT$ particle along with the combined statistical and systematic uncertainties. Figure~\ref{fig:TC2Dlimit} shows the  LSTC cross section limits in a two-dimensional plane  as a function of the $\rhoT$ and $\piT$ masses.

\begin{figure}[htb]
 \centering
 \includegraphics[width=\cmsFigWidth]{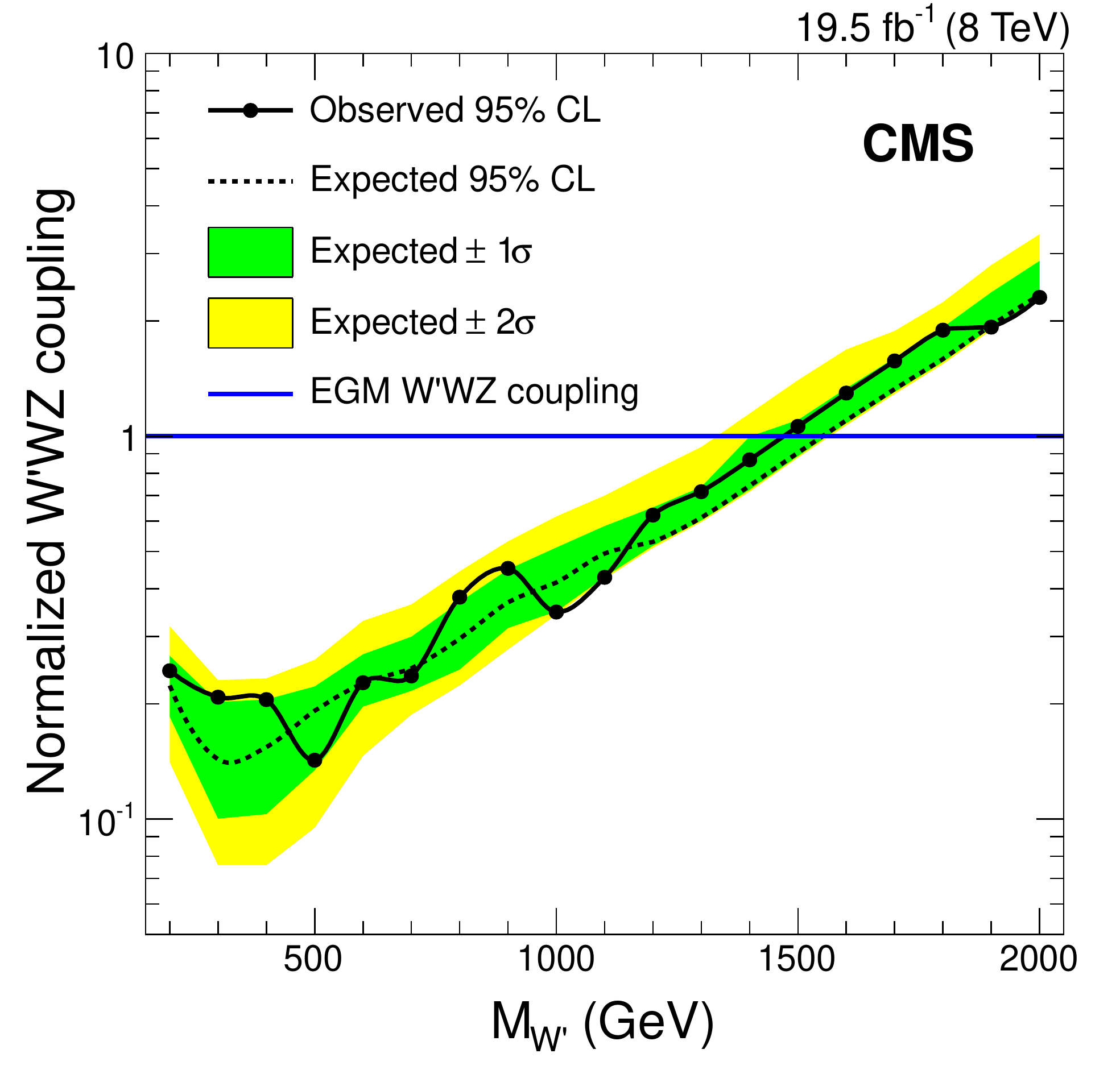}
 \caption{The 95\% CL upper limit on the strength of $\PWpr{\PW\cPZ}$ coupling
normalized to the EGM prediction as a function of the \PWpr\ mass.  The
1\,$\sigma$ and 2\,$\sigma$ combined statistical and systematic uncertainties are indicated
by the green (dark) and  yellow (light) band, respectively. PDF uncertainties on the theoretical cross section are not included.}
 \label{fig:wprimewz_couplinglimit}

 \end{figure}

The \PWpr production cross section and the branching fraction
$\mathcal{B}(\PWpr \to {\PW\cPZ})$ are affected by the strength of the coupling
between the $\PWpr$ boson and $\PW\cPZ$, which we refer to as $g_{\PWpr\PW\cPZ}$.  The EGM assumes that $g_{\PWpr\PW\cPZ}=g_{\PW\PW\cPZ} \times M_{\PW}M_{\cPZ}/M_{\PWpr}^{2}$ where $g_{\PW\PW\cPZ}$ is the SM $\PW\PW\cPZ$ coupling and $M_{\PWpr}$, $M_{\cPZ}$, and $M_{\PW}$ are the masses of the \PWpr, $\cPZ$, and $\PW$ particles, respectively.
If the coupling between the \PWpr\ boson and $\PW\cPZ$ happens to be stronger than that predicted by the EGM, the observed and expected limits will be more stringent. This is
illustrated in Fig.~\ref{fig:wprimewz_couplinglimit}, where an upper limit at 95\% CL on the
$\PWpr{\PW\cPZ}$ coupling is given as a function of the mass of the
$\PWpr$ resonance.

\section{Summary}

A search has been performed in proton-proton collisions at $\sqrt{s}=8$\TeV for new particles decaying via $\PW\cPZ$ to fully leptonic final states with electrons, muons, and neutrinos. The data set corresponds to an integrated luminosity of 19.5\fbinv. No significant excess is found in the mass distribution of the $\PW\cPZ$ candidates compared to the background
expectation from standard model processes. The results are interpreted in the
context of different theoretical models and stringent lower bounds are set at 95\% confidence level on the masses of hypothetical particles decaying via $\PW\cPZ$ to the fully leptonic final state. Assuming an extended gauge model, an expected (observed) exclusion limit of 1.55\;(1.47)\TeV on the mass of the \PWpr boson is set. Low-scale technicolor $\rhoT$ hadrons with masses below 1.14\TeV are also excluded assuming $M_{\piT}=\frac{3}{4}M_{\rhoT} - 25$\GeV. These exclusion limits represent a large improvement over previously published results obtained in proton-proton collisions with $\sqrt{s}=7$\TeV.

\section*{Acknowledgments}

We congratulate our colleagues in the CERN accelerator departments for the excellent performance of the LHC and thank the technical and administrative staffs at CERN and at other CMS institutes for their contributions to the success of the CMS effort. In addition, we gratefully acknowledge the computing centres and personnel of the Worldwide LHC Computing Grid for delivering so effectively the computing infrastructure essential to our analyses. Finally, we acknowledge the enduring support for the construction and operation of the LHC and the CMS detector provided by the following funding agencies: BMWFW and FWF (Austria); FNRS and FWO (Belgium); CNPq, CAPES, FAPERJ, and FAPESP (Brazil); MES (Bulgaria); CERN; CAS, MoST, and NSFC (China); COLCIENCIAS (Colombia); MSES and CSF (Croatia); RPF (Cyprus); MoER, ERC IUT and ERDF (Estonia); Academy of Finland, MEC, and HIP (Finland); CEA and CNRS/IN2P3 (France); BMBF, DFG, and HGF (Germany); GSRT (Greece); OTKA and NIH (Hungary); DAE and DST (India); IPM (Iran); SFI (Ireland); INFN (Italy); NRF and WCU (Republic of Korea); LAS (Lithuania); MOE and UM (Malaysia); CINVESTAV, CONACYT, SEP, and UASLP-FAI (Mexico); MBIE (New Zealand); PAEC (Pakistan); MSHE and NSC (Poland); FCT (Portugal); JINR (Dubna); MON, RosAtom, RAS and RFBR (Russia); MESTD (Serbia); SEIDI and CPAN (Spain); Swiss Funding Agencies (Switzerland); MST (Taipei); ThEPCenter, IPST, STAR and NSTDA (Thailand); TUBITAK and TAEK (Turkey); NASU and SFFR (Ukraine); STFC (United Kingdom); DOE and NSF (USA).

Individuals have received support from the Marie-Curie programme and the European Research Council and EPLANET (European Union); the Leventis Foundation; the A. P. Sloan Foundation; the Alexander von Humboldt Foundation; the Belgian Federal Science Policy Office; the Fonds pour la Formation \`a la Recherche dans l'Industrie et dans l'Agriculture (FRIA-Belgium); the Agentschap voor Innovatie door Wetenschap en Technologie (IWT-Belgium); the Ministry of Education, Youth and Sports (MEYS) of the Czech Republic; the Council of Science and Industrial Research, India; the HOMING PLUS programme of Foundation for Polish Science, cofinanced from European Union, Regional Development Fund; the Compagnia di San Paolo (Torino); the Consorzio per la Fisica (Trieste); MIUR project 20108T4XTM (Italy); the Thalis and Aristeia programmes cofinanced by EU-ESF and the Greek NSRF; and the National Priorities Research Program by Qatar National Research Fund.

\bibliography{auto_generated}   
\cleardoublepage \appendix\section{The CMS Collaboration \label{app:collab}}\begin{sloppypar}\hyphenpenalty=5000\widowpenalty=500\clubpenalty=5000\input{EXO-12-025-authorlist.tex}\end{sloppypar}
\end{document}

%% file: EXO-12-025-authorlist.tex
\textbf{Yerevan Physics Institute,  Yerevan,  Armenia}\\*[0pt]
V.~Khachatryan, A.M.~Sirunyan, A.~Tumasyan
\vskip\cmsinstskip
\textbf{Institut f\"{u}r Hochenergiephysik der OeAW,  Wien,  Austria}\\*[0pt]
W.~Adam, T.~Bergauer, M.~Dragicevic, J.~Er\"{o}, C.~Fabjan\cmsAuthorMark{1}, M.~Friedl, R.~Fr\"{u}hwirth\cmsAuthorMark{1}, V.M.~Ghete, C.~Hartl, N.~H\"{o}rmann, J.~Hrubec, M.~Jeitler\cmsAuthorMark{1}, W.~Kiesenhofer, V.~Kn\"{u}nz, M.~Krammer\cmsAuthorMark{1}, I.~Kr\"{a}tschmer, D.~Liko, I.~Mikulec, D.~Rabady\cmsAuthorMark{2}, B.~Rahbaran, H.~Rohringer, R.~Sch\"{o}fbeck, J.~Strauss, A.~Taurok, W.~Treberer-Treberspurg, W.~Waltenberger, C.-E.~Wulz\cmsAuthorMark{1}
% ...
\vskip\cmsinstskip
% ...
1:~~Also at Vienna University of Technology, Vienna, Austria\\
2:~~Also at CERN, European Organization for Nuclear Research, Geneva, Switzerland\\
% ...
